\documentstyle[11pt,moriond,epsfig,overcite]{article}

\def\lqcd{\Lambda_{\rm QCD}}
\def\qsh{\hat q^2}
\def\qcut{q_0}

\def\gtrsim{\mathrel{\mathpalette\vereq>}}
\def\vereq#1#2{\lower3pt\vbox{\baselineskip1pt\lineskip1pt
    \ialign{\\$#1\hfill##\hfil\\$\crcr#2\crcr\sim\crcr}}}
\begin{document}
\vspace*{4cm}


\title{On $|V_{ub}|$ from the $\bar B\to X_u\ell\bar\nu$ 
dilepton invariant mass spectrum$^{\,*}$}

\author{Christian W.\ Bauer,$^1$ 
  Zoltan Ligeti,$^2$ and Michael Luke$^{\,1}$}

\address{ \vspace*{4pt}
  $^1$Department of Physics, University of Toronto, \\
    60 St.\ George Street, Toronto, Ontario, Canada M5S 1A7 \\ [3pt]
  $^2$Theory Group, Fermilab, P.O.\ Box 500, Batavia, IL 60510 }

\maketitle\abstracts{The invariant mass spectrum of the lepton pair in
inclusive semileptonic $\bar B\to X_u \ell\bar\nu$ decay yields a model
independent determination of $|V_{ub}|$~\cite{BLL}.  Unlike the lepton energy
and hadronic invariant mass spectra, nonperturbative effects are only important
in the resonance region, and play a parametrically suppressed role when ${\rm
d}\Gamma / {\rm d}q^2$ is integrated over $q^2 > (m_B-m_D)^2$, which is
required to eliminate the $\bar B\to X_c \ell\bar\nu$ background.  We discuss
these backgrounds for $q^2$ slightly below $(m_B-m_D)^2$, and point out that
instead of $q^2 > (m_B-m_D)^2 = 11.6\,{\rm GeV}^2$, the cut can be lowered to
$q^2 \gtrsim 10.5\,{\rm GeV}^2$.  This is important experimentally,
particularly when effects of a finite neutrino reconstruction resolution are
included. }

\footnotetext{${}^*$Talk given by Z.L.\ at the 35th Rencontres de Moriond: QCD
and High Energy Hadronic Interactions, Les Arcs, France, March 18--25, 2000. 
\quad UTPT-00-08, FERMILAB-Conf-00/149-T. }

A precise and model independent determination of the Cabibbo-Kobayashi-Maskawa
(CKM) matrix element $V_{ub}$ is important for testing the Standard Model at
$B$ factories via the comparison of the angles and the sides of the unitarity
triangle.  At the present time the allowed range for $\sin2\beta$ in the SM is
largely controlled by the model dependent theory error in $|V_{ub}|$.  

If it were not for the huge background from decays to charm, it would be
straightforward to determine $|V_{ub}|$.  Inclusive $B$ decay rates can be
computed model independently in a series in $\lqcd/m_b$ and $\alpha_s(m_b)$
using an operator product expansion (OPE)~\cite{CGG,incl,MaWi,Blok}, and the
result may schematically be written as
\begin{equation}\label{schematic}
{\rm d}\Gamma = \pmatrix{ b{\rm ~quark} \cr {\rm decay}\cr } \times 
  \bigg\{ 1 + \frac0{m_b} + \frac{f(\lambda_1,\lambda_2)}{m_b^2} + \ldots
  + \frac{\alpha_s}\pi\, (\ldots) + \frac{\alpha_s^2}{\pi^2}\, (\ldots) 
  + \ldots \bigg\} \,.
\end{equation}
At leading order, the $B$ meson decay rate is equal to the $b$ quark decay
rate.  The leading nonperturbative corrections of order $\lqcd^2 / m_b^2$ are
characterized by two heavy quark effective theory (HQET) matrix elements,
usually called $\lambda_1$ and $\lambda_2$.  These matrix elements also occur
in the expansion of the $B$ and $B^*$ masses in powers of $\lqcd/m_b$,
\begin{equation}\label{massrelation}
m_B = m_b + \bar\Lambda 
  - {\lambda_1 + 3 \lambda_2 \over 2m_b} + \ldots \,, \qquad
m_{B^*} = m_b + \bar\Lambda 
  - {\lambda_1 - \lambda_2 \over 2m_b} + \ldots \,.
\end{equation}
Similar formulae hold for the $D$ and $D^*$ masses.  The parameters
$\bar\Lambda$ and $\lambda_1$ are independent of the heavy $b$ quark mass,
while there is a weak logarithmic scale dependence in $\lambda_2$.  The
measured $B^*-B$ mass splitting fixes $\lambda_2(m_b) = 0.12\,{\rm GeV}^2$,
while $\bar\Lambda$ and $\lambda_1$ (or, equivalently, a short distance $b$
quark mass and $\lambda_1$) may be determined from other physical
quantities\cite{FLS,gremmetal,bsg}.  Thus, a measurement of the total
$B\rightarrow X_u \ell\bar\nu$ rate would provide a $\sim 5\%$ determination of
$|V_{ub}|$\cite{upsexp,burels}.

\begin{figure}[t]
\centerline{\epsfysize=5truecm \epsfbox{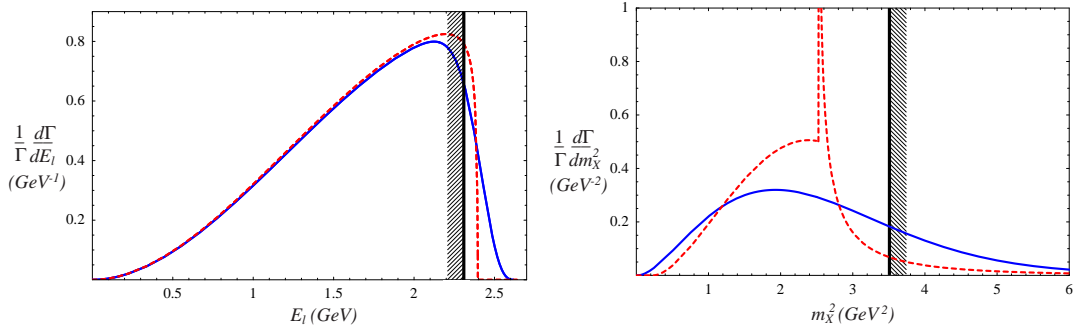} }
\vspace*{-.8cm}
\caption[]{The shapes of the lepton energy and hadronic invariant mass spectra.
The dashed curves are the $b$ quark decay results to ${\cal O}(\alpha_s)$,
while the solid curves are obtained by smearing with the model distribution
function $f(k_+)$ in Eq.~(\ref{sfn}).  The unshaded side of the vertical lines
indicate the region free from charm background.}
\label{twospectra}
\end{figure}

Unfortunately, the $\bar B\to X_u\ell\bar\nu$ rate can only be measured
imposing cuts on the phase space to eliminate the $\sim 100$ times larger $\bar
B\to X_c\ell \bar\nu$ background.  The predictions of the OPE are only model
independent for sufficiently inclusive observables, when hadronic final state
with 
\begin{equation}\label{inequality}
  m_X^2 \gg E_X \lqcd \gg \lqcd^2 
\end{equation}
are allowed to contribute.  Two kinematic regions for which the charm
background is absent have received much attention: the large lepton energy
region,  $E_\ell > (m_B^2-m_D^2)/2m_B$, and the small hadronic invariant mass
region, $m_X < m_D$\cite{FLW,DU}.  However, in both of these regions of phase
space the $\bar B\to X_u\ell\bar\nu$ decay products are dominated by high
energy, low invariant mass hadronic states, for which the inequality
(\ref{inequality}) is violated and the OPE breaks down.  This occurs because
the OPE includes the expansion parameter $E_X\lqcd/m_X^2$ which becomes of
order unity ($m_b\lqcd/m_c^2 \sim 1$ numerically) for $E_X \sim m_b$ and $m_X
\sim m_c$.  To predict the rates in these regions, the complete series in
$E_X\lqcd/m_X^2$ must be resummed into a nonperturbative light-cone
distribution function $f(k_+)$ for the $b$ quark~\cite{shape}.  To leading
order in $1/m_b$, the effects of the distribution function on various 
spectra~\cite{dFN,DU} may be included by replacing $m_b$ by $m_b^* \equiv m_b 
+ k_+$ in the parton level spectrum, ${\rm d}\Gamma_{\rm p}$, and integrating 
over the light-cone momentum
\begin{equation}
{\rm d}\Gamma = \int {\rm d}k_+\, f(k_+)\,
  {\rm d}\Gamma_{\rm p} \Big|_{m_b\to m_b^*} \,.
\end{equation}
The situation is illustrated in Fig.~\ref{twospectra}, where we have
plotted the lepton energy and hadronic invariant mass spectra in the
parton model (dashed curves) and smeared with a simple one-parameter
model for the distribution function (solid curves)~\cite{MN}
\begin{equation}\label{sfn}
f(k_+) = {32\over \pi^2 \Lambda}\, (1-x)^2\, 
  \exp\left[-{4\over \pi}(1-x)^2\right] \Theta(1-x) \,, 
\qquad  x \equiv {k_+\over \Lambda} \,, \qquad \Lambda=0.48\,{\rm GeV}\,.
\end{equation}
While it may be possible to extract $f(k_+)$ from the $B\to X_s\gamma$ photon
spectrum~\cite{shape,LLR}, unknown order $\lqcd/m_b$ corrections are left over,
limiting the accuracy with which $|V_{ub}|$ may be obtained.

\begin{figure}[t]
\centerline{\epsfysize=4.6truecm \epsfbox{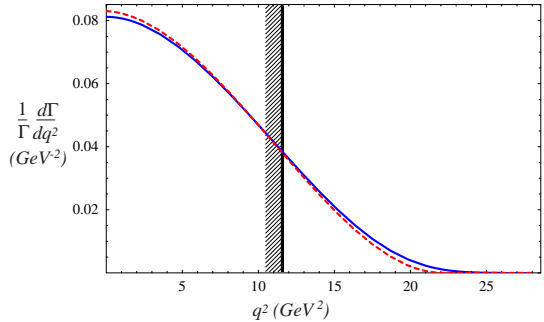} }
\vspace*{-.5cm}
\caption[]{The dilepton invariant mass spectrum.  
The notation is the same as in Fig.~\ref{twospectra}.}
\label{qsqspectrum}
\end{figure}

The situation is very different for the dilepton invariant mass spectrum. 
Decays with $q^2 \equiv (p_\ell + p_{\bar\nu})^2 > (m_B - m_D)^2$ must arise
from $b\to u$ transition.  Such a cut forbids the hadronic final state from
moving fast in the $B$ rest frame, and simultaneously imposes $m_X < m_D$ and
$E_X < m_D$.  Thus, the light-cone expansion which gives rise to the shape
function is not relevant in this region of phase space~\cite{BI,DU}.  This is
also clear from Eq.~(\ref{q2spec}): the contribution of the $\lambda_1$ term to
the decay rate, which is the first term in the shape function, is suppressed
compared to the lowest order term in the OPE for any value of $q^2$.  The
effect of smearing the $q^2$ spectrum with the model distribution function in
Eq.~(\ref{sfn}) is illustrated in Fig.~\ref{qsqspectrum}.  It is clearly a
subleading effect.  The improved behavior of the $q^2$ spectrum over the
$E_\ell$ and $m_X^2$ spectra is also reflected in the perturbation series. 
There are Sudakov double logarithms near the phase space boundaries in the
$E_\ell$ and $m_X^2$ spectra, whereas there are only single logarithms in the
$q^2$ spectrum.

The $\bar B\to X_u\ell\bar\nu$ decay rate with lepton invariant mass above a
given cutoff can be reliably computed working to a fixed order in the OPE
(i.e., ignoring the light-cone distribution function),
\begin{eqnarray}\label{q2spec}
{1\over \Gamma_0}\, {{\rm d}\Gamma \over {\rm d}\qsh} &=&
  \bigg( 1 + {\lambda_1\over 2m_b^2} \bigg)\, 2\, (1-\qsh)^2\, (1+2\qsh) 
  + {\lambda_2\over m_b^2}\, (3 - 45\hat q^4 + 30\hat q^6) \nonumber\\*
&&{} + {\alpha_s(m_b) \over \pi}\, X(\qsh)
  + \bigg( {\alpha_s(m_b) \over \pi} \bigg)^2\, \beta_0\, Y(\qsh) + \ldots \,,
\end{eqnarray}
where $\qsh = q^2/m_b^2$, $\beta_0 = 11 - 2n_f/3$, and $\Gamma_0 = G_F^2\,
|V_{ub}|^2\, m_b^5 / (192\, \pi^3)$ is the tree level $b\to u$ decay rate.  The
ellipses in Eq.~(\ref{q2spec}) denote terms of order $(\lqcd/m_b)^3$ and order
$\alpha_s^2$ terms not enhanced by $\beta_0$.  The function $X(\qsh)$ is known
analytically~\cite{JK}, whereas $Y(\qsh)$ was computed numerically~\cite{LSW}. 
The order $1/m_b^3$ nonperturbative corrections are also known~\cite{m3corr},
as are the leading logarithmic perturbative corrections proportional to
$\alpha_s^n \log^n (m_c/m_b)$~\cite{Matthias}.  The matrix element of the
kinetic energy operator, $\lambda_1$, only enters the $\hat q^2$ spectrum in a
very simple form, because the unit operator and the kinetic energy operator are
related by reparameterization invariance~\cite{LM}.

\begin{figure}[t]
\centerline{\epsfysize=5.4truecm \epsfbox{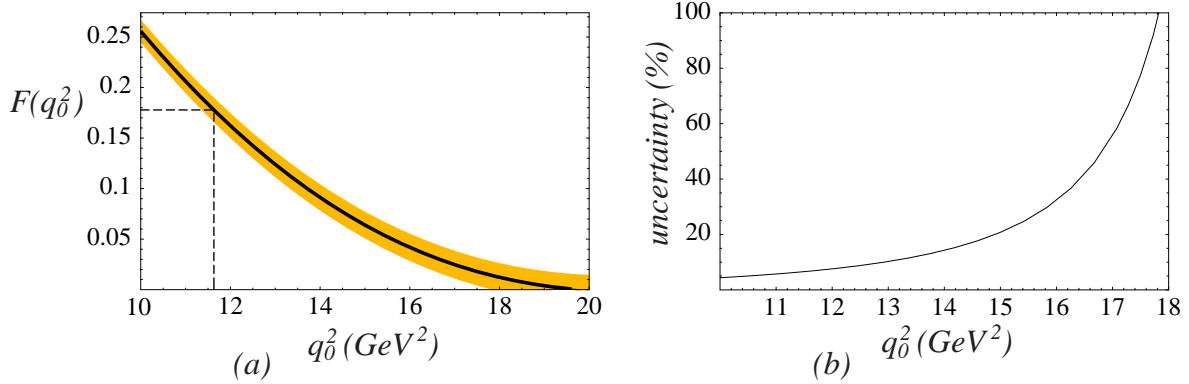} }
\vspace*{-.5cm}
\caption[]{(a) The fraction of $\bar B\to X_u \ell\bar\nu$ events with $q^2 >
\qcut^2$, $F(\qcut^2)$, in the upsilon expansion.  The dashed line indicates 
the lower cut $\qcut^2 = (m_B-m_D)^2 \simeq 11.6\,{\rm GeV}^2$, which 
corresponds to $F = 0.178 \pm 0.012$.  The shaded region is the estimated 
uncertainty due to $\lqcd^3 / m_b^3$ terms; which is shown in (b) as a 
percentage of $F(\qcut^2)$. }
\label{fractionplot}
\end{figure}

The relation between the total $\bar B\to X_u \ell\bar\nu$ decay rate and
$|V_{ub}|$ is known at the $\sim5\%$ level~\cite{upsexp,burels}, 
\begin{equation}\label{Vubups}
|V_{ub}| = (3.04 \pm 0.06 \pm 0.08) \times 10^{-3}\,
  \left( {{\cal B}(\bar B\to X_u \ell\bar\nu)|_{q^2 > \qcut^2} \over 
  0.001 \times F(\qcut^2) }\, {1.6\,{\rm ps}\over\tau_B} \right)^{1/2} \,,
\end{equation}
where $F(\qcut^2)$ is the fraction of $\bar B\to X_u \ell\bar\nu$ events with
$q^2 > \qcut^2$, satisfying $F(0)=1$.  The errors explicitly shown in
Eq.~(\ref{Vubups}) are the estimates of the perturbative and nonperturbative
uncertainties in the upsilon expansion~\cite{upsexp} respectively.  
At the present time the
biggest uncertainty is due to the error of a short distance $b$ quark mass,
whichever way it is defined\cite{Matthias}.  (This can be cast into an 
uncertainty in an
appropriately defined $\bar\Lambda$, or the nonperturbative contribution to the
$\Upsilon(1S)$ mass, etc.)  By the time the $q^2$ spectrum in $\bar B\to X_u
\ell\bar\nu$ is measured, this uncertainty should be reduced from
extracting $m_b$ from the hadron mass~\cite{FLS} or lepton
energy~\cite{gremmetal} spectra in $\bar B\to X_c \ell\bar\nu$, or from the
photon energy spectrum~\cite{bsg} in $B\to X_s\gamma$.  The uncertainty in the
perturbation theory calculation will be largely reduced by computing the full
order $\alpha_s^2$ correction in Eq.~(\ref{Vubups}).  
The largest ``irreducible" uncertainty
is from order $\lqcd^3/m_b^3$ terms in the OPE, the 
estimated size of which is shown in
Fig.~\ref{fractionplot}, together with our central value for $F(\qcut^2)$, as
functions of $\qcut^2$.

\begin{figure}[tb]
\centerline{\epsfysize=5truecm \epsfbox{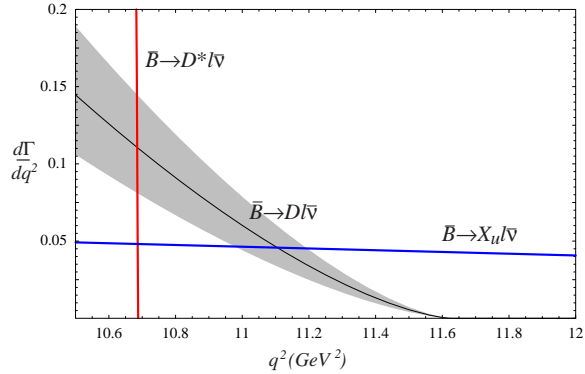} }
\vspace*{-.2cm}
\caption[]{Charm backgrounds near $q^2=(m_B-m_D)^2$.  (Arbitrary units.) }
\label{charmplot}
\end{figure}

There is another advantage of the $q^2$ spectrum over the $m_X$ spectrum to
measure $|V_{ub}|$.  In the variable $m_X$, about 20\% of the charm background
is located right next to the $b\to u$ ``signal region", $m_X < m_D$, namely
$\bar B\to D \ell\bar\nu$ at $m_X = m_D$.  In the variable $q^2$, the charm
background just below $q^2 = (m_B-m_D)^2$ comes from the lowest mass $X_c$
states.  Their $q^2$ distributions are well understood based on heavy quark
symmetry~\cite{HQS}, since this region corresponds to near zero recoil. 
Fig.~\ref{charmplot} shows the $\bar B\to D \ell \bar\nu$ and $\bar B\to D^*
\ell \bar\nu$ decay rates using the measured form factors~\cite{data} (and
$|V_{ub}| = 0.0035$).  The $\bar B\to X_u \ell\bar\nu$ rate is the flat curve. 
Integrated over the region $q^2 > (m_B-m_{D^*})^2 \simeq 10.7\, {\rm GeV}^2$,
the uncertainty of the $B\to D$ background is small due to its $(w^2-1)^{3/2}$
suppression compared to the $\bar B\to X_u\ell\bar\nu$ signal.  This
uncertainty will be further reduced in the near future.  This increases the
$b\to u$ region relevant for measuring $|V_{ub}|$ by $\sim1\, {\rm GeV}^2$. 
The $B\to D^*$ rate is only suppressed by $(w^2-1)^{1/2}$ near zero recoil, and
therefore it is more difficult to subtract it reliably
from the $b\to u$ signal.  The nonresonant $D\pi$ final state
contributes in the same region as $\bar B\to D^*$, and it is reliably predicted to
be small near maximal $q^2$ (zero recoil) based on chiral perturbation
theory~\cite{Dpi}.  The $D^{**}$ states only contribute for $q^2 < 9\, {\rm
GeV}^2$, and some aspects of their $q^2$ spectra are also known model
independently~\cite{Dss}.

Concerning experimental considerations, measuring the $q^2$ spectrum requires
reconstruction of the neutrino four-momentum, just like measuring the hadronic
invariant mass spectrum.  A lepton energy cut may be required for this
technique, however, the constraint $q^2 > (m_B-m_D)^2$ automatically implies
$E_\ell > (m_B-m_D)^2/2m_B \simeq 1.1\,$GeV in the $B$ rest frame.  Even if the
$E_\ell$ cut has to be slightly larger than this, the utility of our method
will not be affected, but a calculation including the effects of
arbitrary $E_\ell$ and $q^2$ cuts would be required.  If experimental resolution
on the reconstruction of the neutrino momentum necessitates a significantly
larger cut than $\qcut^2 = (m_B-m_D)^2$, then the uncertainties in the OPE
calculation of $F(\qcut^2)$ increase.  In this case, it may be possible to
obtain useful model independent information on the $q^2$ spectrum in the region
$q^2 > m_{\psi(2S)}^2 \simeq 13.6\,{\rm GeV}^2$ from the $q^2$ spectrum in the
rare decay $\bar B \to X_s \ell^+\ell^-$, which may be measured in the upcoming
Tevatron Run-II.  

In conclusion, we have shown that the $q^2$ spectrum in inclusive semileptonic
$\bar B \to X_u \ell \bar\nu$ decay gives a model independent determination of
$|V_{ub}|$ with small theoretical uncertainty.  Nonperturbative effects are
only important in the resonance region, and play a parametrically suppressed
role when ${\rm d}\Gamma/{\rm d}q^2$ is integrated over $q^2>(m_B-m_D)^2$,
which is required to eliminate the charm background.  This is a qualitatively
better situation than other extractions of $|V_{ub}|$ from inclusive charmless
semileptonic $B$ decay.

\section*{Acknowledgements}

This work was supported in part by the Natural Sciences and Engineering
Research Council of Canada and the Sloan Foundation.  Fermilab is operated by
Universities Research Association, Inc., under DOE contract DE-AC02-76CH03000.

\end{document}